\documentclass[aps,showpacs,superscriptaddress,twocolumn,amsmath,amssymb,prb]{revtex4}
\usepackage{dcolumn}% Align table columns on decimal point
\usepackage{bm}% bold math
\usepackage[dvips]{graphicx}
\usepackage{wrapfig,subfigure}
\usepackage{amssymb}
\usepackage{color}
\usepackage{ulem}

\newcommand{\ad}{a^{\dagger}}

\begin{document}

\title{Detecting and characterizing frequency fluctuations of vibrational modes}

\author{Z. A. Maizelis}
\affiliation{Department of Physics and Astronomy, Michigan State University, East Lansing, MI 48824}
\author{M. L. Roukes}
\affiliation{Kavli Nanoscience Institute, MC 114-36,
California Institute of Technology, Pasadena, CA 91125}
\author{M. I. Dykman}
\affiliation{Department of Physics and Astronomy, Michigan State University, East Lansing, MI 48824}

\begin{abstract}
We show how frequency fluctuations of a vibrational mode can be separated from other sources of phase noise. The method is based on the analysis of the time dependence of the complex amplitude of forced vibrations. The moments of the complex amplitude sensitively depend on the frequency noise statistics and its power spectrum. The analysis applies to classical and to quantum vibrations.
\end{abstract}
\date{\today}
\maketitle

\section{Introduction}
\label{sec:intro}

In recent years, there has been made significant progress in developing micro- and nano-mechanical systems that display slowly decaying vibrations. For different types of such systems, the ratio of the vibration eigenfrequency to the decay rate, the quality factor $Q$, has reached $\gtrsim 10^5$. \cite{Schwab2005a,Steele2009,Lassagne2009} This has allowed studying new physics, including quantum phenomena,\cite{Chan2001a,Anetsberger2010} and opened a way for numerous applications, like highly-sensitive mass sensing \cite{Jensen2008,Naik2009,Lee2010} and, potentially, high-accuracy nanomechanical clocks. In parallel, high-Q modes of superconducting cavities have been used for control and measurement of Josephson-junction based qubits.\cite{Schuster2005}

An important problem in the studies of nanomechanical vibrations and superconducting cavity modes is to understand the mechanisms of their decay and loss of coherence. Often one separates decay and fluctuations of the vibration amplitude and fluctuations of the vibration phase. Phase fluctuations are not only interesting on their own but are particularly important for applications, as they can impose limits on the sensitivity of a device. They can come from the thermal noise that accompanies vibration decay and is a consequence of coupling to a thermal reservoir. A more delicate and often more important source is fluctuations of the vibration frequency. They can have various origins, see Ref.~\onlinecite{VanKampen1976} and papers cited therein, with recent examples being random attachment or detachment of molecules to a resonator that changes its mass, \cite{Yong1989,*Yong1990, Naik2009,Lee2010,Dykman2010} molecule diffusion along the resonator, \cite{Atalaya2011,Yang2011} coupling of the vibrational mode to two-state fluctuators, \cite{Gao2008} and, for nonlinear vibrations, frequency modulation by fluctuations of the vibration amplitude. \cite{DK_review84}

In this paper, we suggest a simple way of separating and characterizing frequency fluctuations in vibrational systems.
In two-level systems, frequency fluctuations lead to the difference between  the $T_1$ and $T_2$ relaxation times and are routinely separated from decay using nonlinear response to an external field. \cite{Slichter1990} In contrast, the response of linear vibrations is inherently linear, and the spectrum of the response remains a major source of information about the dynamics. If frequency fluctuations are the dominating factor, this spectrum reveals some of their features.\cite{Kubo1963,VanKampen1976,Yong1989,Dykman2010} However, in many cases of interest it does not provide enough information, and often does not allow one to even detect frequency fluctuations at all. For example, for broadband Gaussian frequency noise, the absorption spectrum is Lorentzian, as if there were no frequency noise, even though the overall width of the spectrum exceeds the width due to decay.

We show below that frequency fluctuations can be studied by using, in a different way, essentially the same measurement as that used to find the absorption spectrum, i.e., by looking at the response of a resonantly modulated oscillator. This applies to both classical and quantum oscillators. The idea is to study such correlators of the quadrature and in-phase components of the oscillator displacement that are specifically sensitive to frequency fluctuations. As we show, these are correlators and moments of the complex vibration amplitude. They allow one not only to reveal frequency noise, but also to study its statistics, for both classical and quantum vibrations. The sensitivity to the noise statistics is illustrated for  important examples of the noise.

\section{Equation of motion for the quadratures}

We assume that the oscillator energy relaxation comes from the coupling to a thermal reservoir, which is linear in the oscillator coordinate and momentum and weak, so that the decay rate $\Gamma \ll \omega_0$, where $\omega_0$ is the oscillator eigenfrequency in the absence of frequency noise. With the noise the frequency becomes $\omega_0+\xi(t)$. We assume that the  frequency noise $\xi(t)$ is a stationary process and $\langle \xi(t)\rangle = 0$. Of primary interest is the case of small noise,where overall frequency fluctuations are small compared to $\omega_0$.

Phenomenologically, the motion of the oscillator with coordinate $q$ and with unit mass in the presence of a driving force $F\cos\omega_Ft$  is described by equation
\begin{eqnarray}
\label{eq:eom_phenomenol}
\ddot q + 2\Gamma \dot q + [\omega_0^2+ 2\omega_0\xi(t)]q = F\cos\omega_Ft + f(t),
\end{eqnarray}
where $f(t)$ is zero-mean additive thermal noise. We will be considering this motion in the rotating frame on  times much longer than $\omega_0^{-1}$. On this scale, the approximation of Markovian relaxation of the oscillator amplitude and phase applies even where Eq.~(\ref{eq:eom_phenomenol}) does not apply; in other words, the assumption of Ohmic dissipation is not required. \cite{Schwinger1961}

We assume that the typical frequencies of the noise $\xi(t)$ are small compared to $\omega_0$.  This is the case for many systems  of current interest; coupling of the oscillator to the source of such noise does not lead to energy relaxation via nonlinear friction\cite{Dykman1975a} and to random parametric excitation of the oscillator. \cite{Lindenberg1981,Gitterman_book2005}

We consider resonant driving, with frequency $\omega_F$ close to $\omega_0$, i.e., $|\delta\omega|\ll \omega_0$, where $\delta\omega=\omega_F-\omega_0$.
The oscillator dynamics can be then conveniently analyzed by changing from the coordinate $q$ and momentum $p=\dot q$ to slowly varying on the time scale $\omega_F^{-1}$ complex variables $u(t), u^*(t)$,
\begin{eqnarray}
\label{eq:slow_variables}
&&q(t)=u\exp(i\omega_Ft) + u^*\exp(-i\omega_Ft),\\
&&p(t)= i\omega_F\left[u\exp(i\omega_Ft) - u^*\exp(-i\omega_Ft)\right].\nonumber
\end{eqnarray}
Function $u(t)$ is the complex vibration amplitude. From Eq.~(\ref{eq:slow_variables}), Re~$u$ and Im~$u$ give, respectively, the in-phase and  quadrature components of the oscillator displacement at frequency $\omega_F$.

On the time scale that largely exceeds the correlation time of the thermal reservoir and $\omega_F^{-1}$, both a microscopic theory for weak oscillator-to-reservoir coupling \cite{Schwinger1961,Mandel1995,DK_review84} and the phenomenological model of Eq.~(\ref{eq:eom_phenomenol})  lead to an equation of motion for $u$ of the form
\begin{eqnarray}
\label{eq:eom_slow_time}
\dot u \approx -[\Gamma + i\delta\omega - i\xi(t)] u -\frac{iF}{4\omega_F}+f_u(t),
\end{eqnarray}
where $f_u(t)=-(i/2\omega_F) f(t)\exp(-i\omega_Ft)$; the renormalization of $\omega_0$ due to the coupling to a thermal reservoir that emerges in the microscopic theory has been incorporated into $\omega_0$.

For large times compared to the decay time $\Gamma^{-1}$, the initial state of the oscillator is ``forgotten'' and $u(t)$ becomes a linear superposition of the terms that describe forced oscillator vibrations and thermal fluctuations,
\begin{eqnarray}
\label{eq:sol_for_u}
&&u(t)=(F/4\omega_F)u_F(t)+ u_{\rm th}(t),\\
&&u_F(t)=\int_{-\infty}^tdt_1\chi^*(t-t_1)
\exp\left[i\int_{t_1}^tdt_1'\xi(t_1')\right],\nonumber\\
&&\chi(t)=i\exp(-\Lambda^* t),\qquad \Lambda\equiv\Lambda(\omega_F) =\Gamma + i\delta\omega.\nonumber
\end{eqnarray}
Here, $u_F(t)$ is the scaled complex amplitude of  forced vibrations; $\chi(t)$ is the scaled oscillator susceptibility in the absence of frequency fluctuations. The term $u_{\rm th}$ comes from the additive thermal noise,
\begin{equation}
\label{eq:random_u}
u_{\rm th}(t)=i\int_{-\infty}^tdt_1\chi^*(t-t_1)f_u(t_1)%\nonumber\\
%&&\times
\exp\left[i\int_{t_1}^tdt_1'\xi(t_1')\right].
\end{equation}

We are interested in the effects of frequency noise, not additive noise. By increasing the field $F$, the term $\propto u_F(t)$ in $u(t)$, Eq.~(\ref{eq:sol_for_u}), can be made larger than the typical value of $u_{\rm th}$. However, as we will show, the contribution of additive noise to the correlators of $u(t)$ vanishes in the approximation used to derive Eq.~(\ref{eq:eom_slow_time}), which allows one to use even comparatively weak driving fields for studying frequency noise.

\section{Correlators of the complex amplitude: Independence of additive noise}

The two noises that determine the oscillator dynamics, $f(t)$ and $\xi(t)$, are uncorrelated, generally. The noise $f(t)$ results from the linear in $q,p$ coupling to a thermal reservoir. The major effect on the oscillator comes from the Fourier components of $f(t)$ with frequencies $\omega$ such that $|\omega-\omega_0|\ll \omega_0$. It was assumed in deriving the Markovian equation of motion for $u(t)$, Eq.~(\ref{eq:eom_slow_time}), that the spectral density of $f(t)$
is smooth around $\omega_0$ and can be set equal to a constant  for $|\omega-\omega_0|\lesssim |\delta\omega|, \langle\xi^2(t)\rangle^{1/2}$.\cite{Schwinger1961,Mandel1995,DK_review84}

In contrast, noise $\xi(t)$ comes from either an external nonequilibrium source or from the interaction with a thermal reservoir that is effectively quadratic in $q,p$ and couples the oscillator to the degrees of freedom other than those that lead to $f(t)$.

An important consequence of the statistical independence of $f(t)$ and $\xi(t)$ is that
\begin{equation}
\label{eq:random_term_average}
\langle u_{\rm th}^n(t)\rangle =0,\qquad n=1,2,3,\ldots .
\end{equation}
A simple way to see this is by comparing the expressions for $u_{\rm th}(t)$ and $u_{\rm th}(t+t_0)$ with an arbitrary $t_0$. If one writes Eq.~(\ref{eq:random_u}) for $u_{\rm th}(t+t_0)$ and changes from integrating over $t_1$ and $t_1'$ to $\tilde t_1 = t_1-t_0$ and $\tilde t_1'=t_1'-t_0$, respectively, the expression for $u_{\rm th}(t+t_0)$ becomes of the same form as $u_{\rm th}(t)$, except that the noises $f(t)$ and $\xi(t)$ are evaluated for the time shifted by $t_0$ and there emerges an extra factor $\exp(-i\omega_Ft_0)$ from the interrelation between $f_u(t)$ and $f(t)$. For stationary noises $f(t)$ and $\xi(t)$, the change of the origin of time does not affect any average values, and therefore the only difference between averaging $u_{\rm th}(t)$ and $u_{\rm th}(t+t_0)$ is the factor $\exp(-i\omega_Ft_0)$. The average moments of $u_{\rm th}(t)$ may not depend on $t_0$, and therefore they are equal to zero. So are also correlators $\langle u_{\rm th}(t_1)\ldots u_{\rm th}(t_n)\rangle$.

From the above arguments,
\begin{equation}
\label{eq:relevant_moments_}
\langle u(t_1)\ldots u(t_n)\rangle = (F/4\omega_F)^n\langle u_F(t_1)\ldots u_F(t_n)\rangle,
\end{equation}
and below we will be interested in calculating correlators of $u_F$. They are independent from additive noise, for a linear oscillator, and therefore measuring them immediately reveals frequency noise. We note that the $n$th-order correlator is proportional to the $n$th power of the driving field; still the oscillator response remains linear.

\section{Quasiwhite frequency noise}

The explicit expressions (\ref{eq:sol_for_u}) and (\ref{eq:relevant_moments_}) allow one to analyze correlators of the complex amplitude for various types of frequency noise. We will discuss several noise models of interest for experiment and show how by measuring the correlators one can study the noise statistics.

One of the most important is noise with a comparatively broad frequency spectrum, which is flat up to a characteristic cutoff frequency $\omega_{\rm corr}$ such that $\Gamma, |\delta\omega| \ll \omega_{\rm corr}\ll \omega_0$. Such noise is effectively $\delta$ correlated on a time scale long compared to $\omega_0^{-1}$. It can come, for example, from quasielastic scattering of phonons or other excitations off the oscillator,\cite{Ivanov1965,Elliott1965} in which case it is approximately Gaussian, or from the discreteness of the electric current that modulates the oscillator, in which case it is close to Poissonian, or it can come from other processes and have a different statistics. For a $\delta$-correlated noise, it is convenient to do the averaging in Eq.~(\ref{eq:relevant_moments_}) using the noise characteristic functional, which can be written as
\begin{eqnarray}
\label{eq:charact_fnctnl}
{\cal P}[k(t)]&\equiv &\left\langle \exp\left[i\int dt k(t)\xi(t)\right]\right\rangle\nonumber\\
&& = \exp\left[-\int dt \mu\bigl(k(t)\bigr)\right].
\end{eqnarray}
Function $\mu(k)$ is determined by the noise statistics;  for zero-mean Gaussian noise of intensity $D$, with $\langle\xi(t)\xi(t')\rangle=2D\delta(t-t')$, and for zero-mean Poisson noise $\xi(t)=g\sum_n\delta (t-t_n) - g\nu$ with pulse area $g$ and pulse rate $\nu$ we have, respectively, $\mu= \mu_G$ and $\mu=\mu_P$, where \cite{FeynmanQM}
\begin{eqnarray}
\label{eq:Gauss_Poisson}
\mu_G(k)=Dk^2,\qquad \mu_P(k)= \nu\left(1-e^{ikg} + ikg\right).
\end{eqnarray}

For $\delta$-correlated frequency noise, the response of the oscillator to the driving as a function of frequency detuning $\delta\omega=\omega_F-\omega_0$ is of the same functional form as without the noise. From Eqs.~(\ref{eq:sol_for_u}) and (\ref{eq:charact_fnctnl}),
\begin{eqnarray}
\label{eq:Lorentzian_response}
&&\langle u\rangle = (F/4\omega_F)\langle u_F(t)\rangle = -i(F/4\omega_F)(\tilde\Gamma + i\delta\tilde{\omega})^{-1},\nonumber\\
&&\tilde\Gamma= \Gamma + {\rm Re}~\mu(1),\qquad \delta\tilde{\omega}=\delta\omega + {\rm Im}~\mu(1).
\end{eqnarray}
The noise leads to broadening of the spectrum of the response and, generally, to the shift of the oscillator frequency. Both are determined by the value of function $\mu$ for $k=1$. In particular, for quasiwhite noise, the increment of the half-width is given by the noise intensity $D$, a well-known result, whereas for Poisson noise this increment is $\nu (1-\cos g)$, it oscillates with increasing $g$ and increases with the pulse rate $\nu$.

It follows from Eq.~(\ref{eq:Lorentzian_response}) that, from the oscillator spectrum taken alone, one cannot tell whether there is frequency noise at all. However, the pair correlator of the complex amplitude makes it possible to identify the presence of the noise. A straightforward but somewhat lengthy calculation shows that
\begin{eqnarray}
\label{eq:delta_corr_2nd_correlator}
\langle u(t)u(0)\rangle &-&\langle u\rangle^2=\langle u\rangle^2\frac{2\mu(1)-\mu(2)}{2\Lambda + \mu(2)}\nonumber\\
&&\times \exp\left[-(\tilde\Gamma + i\delta\tilde\omega)t\right] \qquad (t>0).
\end{eqnarray}

Because of the frequency noise, $\langle u^2\rangle \neq \langle u\rangle^2$. From Eq.~(\ref{eq:delta_corr_2nd_correlator}), the variance of the complex amplitude $u(t)$ is $\propto 2\mu(1)-\mu(2)$, and is thus determined by the nonlinearity of the function $\mu(k)$. In particular, for Gaussian and Poisson noises we have, respectively, $2\mu_G(1)-\mu_G(2)=-2D$ and $2\mu_P(1)-\mu_P(2) = \nu[1-\exp(ig)]^2$. The time decay of the pair correlator of $\delta u(t)=u(t)-\langle u \rangle$ is exponential, with exponent $\Lambda + \mu(1)\equiv\tilde\Gamma + i\delta\tilde\omega$.

Not only does the pair correlator, Eq~(\ref{eq:delta_corr_2nd_correlator}), allow one to reveal frequency noise where there are no conventional spectral signatures of it, but it also gives an insight into the noise statistics. More insights can be gained from the higher-order moments of $u(t)$. By writing
\begin{eqnarray*}
\label{eq:ordered_moments}
&&u_F^n(0) = n!\int_{-\infty}^0dt_1\int_{-\infty}^{t_1}dt_2\ldots \int_{-\infty}^{t_{n-1}}dt_n (-i)^n\nonumber\\
&&\times\exp\left\{\sum_{j=1}^{n}\left[\Lambda t_j+ i(n+1-j)\int_{t_{j}}^{t_{j-1}}dt'_j\xi(t'_j)\right]\right\},
\end{eqnarray*}
we obtain from Eq.~(\ref{eq:charact_fnctnl})
\begin{eqnarray}
\label{eq:moments_delta_corr}
\langle u^n\rangle = n!\left(\frac{-iF}{4\omega_F}\right)^n\prod_{j=1}^n\left[j\Lambda + \mu(j)\right]^{-1}.
\end{eqnarray}
From Eq.~(\ref{eq:moments_delta_corr}), by measuring the moments of the complex amplitude $u(t)$, one can find function $\mu(k)$ for all integer $k$ and therefore, given that this function is analytical at least near the real-$k$ axis, find the whole $\mu(k)$ and thus the full statistics of the $\delta$-correlated noise $\xi(t)$.

\section{Comparatively weak frequency noise}

The presence of non-$\delta$-correlated frequency noise can be usually directly seen in the spectrum of the oscillator response, if the noise is sufficiently strong. For example, the absorption spectrum may have a fine structure or become asymmetric. \cite{VanKampen1976,Naik2009,Lee2010,Yong1989,Dykman2010,Atalaya2011} The situation is more complicated where the noise is comparatively weak, so that the shape of the spectrum is weakly distorted compared to the Lorentzian contour. We now show that the moments of the complex amplitude allow one to detect frequency noise and study its statistics even in this case.

We will express the moments in terms of the correlators of the frequency noise. The lowest-order correlators of a stationary zero-mean noise are
\begin{eqnarray}
\label{eq:noise_correlators}
&&\Xi_2(\omega)=(2\pi)^{-1}\int dt e^{i\omega t}\langle \xi(t)\xi(0)\rangle,\\
&&\Xi_3(\omega_1,\omega_2)=(2\pi)^{-2}\int dt_1 dt_2 e^{i(\omega_1t_1 + \omega_2t_2)}
\langle \xi(t_1)\xi(t_2)\xi(0)\rangle .\nonumber
\end{eqnarray}
Since $\xi(t_i)$ for different $t_i$ commute with each other, we have 
\begin{eqnarray}
\label{eq:symmetry_properties}
&&\Xi_2(\omega)=\Xi_2(-\omega), \qquad \Xi_3(\omega_1,\omega_2)=\Xi_3(\omega_2,\omega_1)\nonumber\\
&&\qquad=\Xi_3(-\omega_1-\omega_2,\omega_2) = \Xi_3(\omega_1,-\omega_1-\omega_2).
\end{eqnarray}
By expanding Eq.~(\ref{eq:sol_for_u}) for $u_F$ to third order in $\xi(t)$, we obtain
%
%\begin{widetext}
%
\begin{eqnarray}
\label{eq:3rd_order_linear_u}
&&\frac{\langle u\rangle}{\langle u\rangle^{(0)}} \approx 1 -\int \frac{d\omega}{\Lambda^2+\omega^2}\Xi_2(\omega)
-i\int d\omega_1\,d\omega_2\Xi_3^{(\Lambda)}(\omega_1,\omega_2)\nonumber\\
&&\quad\times \Lambda\left(3\Lambda^2 + \omega_1^2 + \omega_1\omega_2 + \omega_2^2\right)/3,
\end{eqnarray}
where $\langle u\rangle^{(0)}=-iF/(4\omega_F\Lambda)$ is the complex amplitude that describes forced vibrations in the absence of frequency noise and
\begin{eqnarray}
\label{eq:xi_Lambda}
&&\Xi_3^{(\Lambda)}(\omega_1,\omega_2)\nonumber\\
&&=\frac{\Xi_3(\omega_1,\omega_2)}{
(\Lambda^2+\omega_1^2)(\Lambda^2+\omega_2^2)[\Lambda^2+(\omega_1+\omega_2)^2]}.
\end{eqnarray}
%
%\end{widetext}
%
It is clear from Eq.~(\ref{eq:3rd_order_linear_u}) that, to third order in $\xi(t)$, the effect of frequency noise on the spectrum of the oscillator response can be described as renormalization of the decay rate $\Gamma$ and the eigenfrequency $\omega_0$, and therefore from spectroscopic data it is hard to tell whether weak frequency noise is present at all.

Frequency noise can be detected by measuring higher moments of the complex amplitude. Keeping only the second and third-order correlators of $\xi(t)$, we obtain for the variance of the complex amplitude
\begin{eqnarray}
\label{eq:2nd_moment_weak_noise}
&&\frac{\langle u^2\rangle - \langle u\rangle^2}{\langle u\rangle^2}
\approx -\int \frac{d\omega}{\Lambda^2 + \omega^2} \Xi_2(\omega)\nonumber\\
&&\quad-i\int d\omega_1d\omega_2\Xi_3^{(\Lambda)}(\omega_1,\omega_2)
\nonumber\\
&&\quad \times\left[2\Lambda^3 + \Lambda(\omega_1^2 + \omega_2^2) + i\omega_1\omega_2(\omega_1+\omega_2)\right].
\end{eqnarray}
Here, we have used the symmetry properties of $\Xi_2$ and $\Xi_3$, Eq.~(\ref{eq:symmetry_properties}); note that if the noise $\xi(t)$ has time-reversal symmetry, the term $\propto \omega_1\omega_2(\omega_1+\omega_2)$ in the integrand in Eq.~(\ref{eq:2nd_moment_weak_noise}) can be disregarded.

For weak noise $\xi(t)$, the leading contribution to the variance of $u$ comes from the second-order term $\propto \Xi_2$. To reveal a nonzero third-order noise correlator, in addition to the variance of $u$ one should measure the third cumulant of $u$,
\begin{eqnarray}
\label{eq:3rd_u_cumulant}
&&\frac{\langle u^3\rangle - 3\langle u\rangle\langle u^2\rangle +2\langle u\rangle^3}{\langle u\rangle^3}
\approx -i\int d\omega_1d\omega_2\Xi_3^{(\Lambda)}(\omega_1,\omega_2)
\nonumber\\
&& \times\left[\Lambda^3 + \Lambda(\omega_1^2 + \omega_2^2 + \omega_1\omega_2) + i\omega_1\omega_2(\omega_1+\omega_2)\right].
\end{eqnarray}
One can see that, in the case of weak $\delta$-correlated noise, Eqs.~(\ref{eq:Lorentzian_response}) --- (\ref{eq:moments_delta_corr}) agree with Eqs.~(\ref{eq:3rd_order_linear_u}) --- (\ref{eq:3rd_u_cumulant}). However, the results of this section are not limited to $\delta$-correlated noise.

\section{Quantum formulation}

The above arguments can be immediately extended to the quantum regime, since the responses of quantum and classical harmonic oscillators to resonant modulation are the same. In the absence of coupling to a thermal reservoir, the Hamiltonian of the oscillator in a resonant field $F\cos\omega_Ft$ in the presence of weak classical frequency noise $\xi(t)$ is
\begin{equation}
\label{eq:H_0}
H_0=\hbar [\omega_0+\xi(t)]a^{\dagger}a - qF\cos\omega_Ft,
\end{equation}
where $a= (2\hbar\omega_0)^{-1/2}(\omega_0q + ip)$ is the lowering operator of the oscillator.

The effect of coupling to a thermal reservoir can be conveniently analyzed in the rotating wave approximation by going to the interaction representation with operator $U(t)=\exp(-i\omega_F a^{\dagger}at)$. When typical noise frequencies are small compared to $\omega_F$, the resulting equation for the oscillator density matrix $\rho_0$, for a given realization of the frequency noise, has a familiar form
\begin{eqnarray}
\label{eq:master_general}
\dot \rho_0 &=& i[\delta\omega -\xi(t)][a^{\dagger}a,\rho_0] -\hat\Gamma\rho_0\nonumber\\
&&+i[F'a^{\dagger}+F'^*a,\rho_0].
\end{eqnarray}
Here, $F'=(8\hbar\omega_0)^{-1/2}F$, operator $\hat\Gamma$ describes oscillator decay, $\hat\Gamma\rho=\Gamma(\bar n+1)(\ad a\rho-2a\rho\ad + \rho\ad a)+\Gamma\bar n (a\ad\rho-2\ad\rho a + \rho a\ad)$, where $\bar n= [\exp(\hbar\omega_0/k_BT)-1]^{-1}$ is the Planck number; as in the classical analysis, the renormalization of the oscillator frequency due to the coupling to the thermal reservoir is assumed to have been incorporated into $\omega_0$.
We emphasize that the density matrix $\rho_0$ has not been averaged over the realizations of $\xi(t)$, it fluctuates in time.

Equation (\ref{eq:master_general}) allows one to perform averaging over thermal fluctuations of the oscillator for given $\xi(t)$. It leads to a chain of equations for the moments $\overline{a^n}(t)\equiv {\rm Tr}[a^n\rho_0(t)]$ of the operator $a$,
\begin{eqnarray}
\label{eq:chain_for_moments}
\frac{d}{dt}\overline{a^n}= -n[\Lambda^*+i\xi(t)]\overline{a^n}+inF'\overline{a^{n-1}}
\end{eqnarray}
with $\Lambda = \Gamma + i\delta\omega$, cf. Eq.~(\ref{eq:sol_for_u}). If we assume that the field $F$ is turned on adiabatically at $t\to -\infty$,  the initial condition to Eq.~(\ref{eq:chain_for_moments})  is $\overline{a^n}\to 0$ for $t \to -\infty$.

The solution of Eq.~(\ref{eq:chain_for_moments}) is
\begin{eqnarray}
\label{eq:solution_for_a}
\overline{a^n}(t)=\left[\overline{a}(t)\right]^n, \qquad \overline{a}(t)=F'u^*_F(t),
\end{eqnarray}
where $u_F(t)$ is given by Eq.~(\ref{eq:sol_for_u}). Therefore the averaging of the moments of operator $a$ over realizations of $\xi(t)$, that gives the mean value $\langle a^n(t)\rangle$, can be done in the same way as for a classical oscillator. Hence, the results for a linear classical oscillator immediately apply to a linear quantum oscillator.

\section{Markov frequency noise}

The analysis of the effect of frequency noise is simplified in the case where the noise is Markovian. Such noise can be continuous, as in the case of frequency fluctuations due to diffusion of massive particles along a nanoresonator, \cite{Lee2010,Yang2011,Atalaya2011} or discrete (i.e., takes on discrete values), as in the case of random attachment or detachment of massive particles to a mechanical resonator \cite{Yong1989,Naik2009,Dykman2010} or transitions between quantum states of a nonlinearly coupled vibrational mode of a trapped electron. \cite{Peil1999} In both cases, its probability distribution $p(\xi,t)$ is described by equation $\dot p = \hat Wp$. Operator $\hat W$ is independent of time, for a stationary process. For a continuous process $\xi(t)$, $\hat W$ is a differential operator with respect to $\xi$; for example, for a diffusion process $\xi(t)$, the equation for $p$ is the Fokker-Planck equation.\cite{VanKampen1976} For a discrete process, $\hat W$ describes transitions between different discrete values of $\xi$, with appropriate transition rates.

Instead of the density matrix $\rho_0$, which depends on a realization of the noise $\xi(t)$, for Markovian $\xi(t)$, it is convenient to introduce density matrix $\rho(\xi,t)$, which remains an operator with respect to the oscillator variables, but also depends on $\xi$ as a variable. For continuous $\xi(t)$, we have $\rho(\xi,t)=\langle \rho_0(t)\delta\bigl(\xi-\xi(t)\bigr)\rangle_{\xi}$, where $\langle\ldots\rangle_{\xi}$ means averaging over realizations of $\xi(t)$. For discrete $\xi(t)$, one should use the same definition, but with Kronecker's delta instead of the $\delta$ function.

The equation for $\rho(\xi,t)$ is an obvious extension of Eq.~(\ref{eq:master_general}),
\begin{eqnarray}
\label{eq:master_over_noise}
\dot\rho &=& i[\delta\omega -\xi][a^{\dagger}a,\rho] - \hat\Gamma\rho + \hat W\rho\nonumber\\
&&+i[F'a^{\dagger}+F'^*a,\rho], \qquad \rho\equiv \rho(\xi,t).
\end{eqnarray}
In this model, there is no backaction from the oscillator on the frequency noise source, and therefore operator $\hat W$ does not depend on the dynamical variables of the oscillator. Then one can immediately write a system of equations for the moments $A(n,\xi,t)={\rm Tr}\left[ a^n\rho(\xi,t)\right]$ of operator $a$, which will now be functions of the random variable $\xi$ rather than functionals of $\xi(t)$,
\begin{eqnarray}
\label{eq:chain_for_xi_functions}
&&\partial_t A(n,\xi,t) = -n[\Lambda^*+i\xi]A(n,\xi,t)+inF'A(n-1,\xi,t)\nonumber\\
 &&\qquad + \hat W A(n,\xi,t),
%\\ &&
\qquad\langle a^n(t)\rangle = \int d\xi A(n,\xi,t)
\end{eqnarray}
(for discrete-valued noise, the integral over $d\xi$ should be replaced by a sum).

For $n=1$, the stationary solution of Eq.~(\ref{eq:chain_for_xi_functions})  was discussed earlier for several models of frequency noise.\cite{Dykman2010,Atalaya2011} As a function of detuning $\delta\omega$, $\langle a\rangle$ gives the spectrum of the response of the oscillator to a resonant force. As mentioned above, where the noise $\xi(t)$ is strong compared to $\Gamma$ (but still weak compared to $\omega_0$), it can significantly change the spectrum compared to the $\xi=0$ case, making it possible to detect the presence of the noise and find some of its characteristics.

By studying the moments of the complex amplitude $\langle a^n\rangle$ one can extract much more information about the frequency noise than just from the spectrum. We note that one can think of functions $A(n,\xi,t)$ in the stationary regime as ``partial moments" of the oscillator for a given eigenfrequency $\omega_0+\xi$.  Equation~(\ref{eq:chain_for_xi_functions}) shows that functions $A(n,\xi,t)$ with the same $n$ but different $\xi$ are coupled by the operator $\hat W$. This is a direct analog of the effect of the interference of the oscillator partial spectra, which was discussed earlier.\cite{DK_review84,Dykman2010}

We will consider as an example the moments $\langle a^n\rangle$ for telegraph noise. This noise takes on two values $\xi_k$  ($k=1,2$), between which it switches at random at rates $W_{12}$ and $W_{21}$. Respectively, in the stationary regime, $A(n,\xi)\equiv A(n,\xi,t)$ has two components, $A(n,\xi_1)$ and $A(n,\xi_2)$, which can be considered as components of a vector ${\bf A}(n)$; operator $\hat W$ becomes a $2\times 2$ matrix, and Eq.~(\ref{eq:chain_for_xi_functions}) can be written as
\begin{eqnarray}
\label{eq:telegraph_matrix}
&&\hat M(n){\bf A}(n)=inF'{\bf A}(n-1),\\
&&\hat M(n)=n(\Lambda^* + i\bar\xi)\hat I +in\xi_c\hat\sigma_z + \hat W,\nonumber
\end{eqnarray}
where $\hat I$ and $\hat\sigma_z$ are the unit matrix and the Pauli matrix, respectively, and
\[\hat W=\left(\begin{array}{cc}W_{12}& -W_{21}\\
-W_{12} & W_{21}
\end{array}\right), \qquad \bar\xi=(\xi_1 + \xi_2)/2;\]
$\xi_c=(\xi_1-\xi_2)/2$ is the amplitude of the frequency noise.

Equation (\ref{eq:telegraph_matrix}) has a simple solution:
\begin{equation}
\label{eq:telegraph_solution}
{\bf A}(n)=(iF')^n\prod_{k=n}^1\hat M^{-1}(k){\bf A}(0),
\end{equation}
where ${\bf A}(0)=(W_{21}/W,W_{12}/W)$, with $W=W_{12}+W_{21}$ being the total switching rate.

For the noise amplitude $|\xi_c|\gg W,\Gamma$ the absorption spectrum of the oscillator, which is given by Im~$\left[A(1,\xi_1)+A(1,\xi_2)\right]$, has two distinct peaks. On the other hand, for $|\xi_c|\lesssim \max{\Gamma, W}$ the peaks are not resolved and the spectrum is a single-peak curve,\cite{Anderson1954,VanKampen1976} which makes it complicated to identify the presence of the frequency noise. The moments of the complex amplitude are advantageous in this respect, as discussed in the next section.

\section{Comparing different types of frequency noise}

\begin{figure}[h]
\includegraphics[scale=0.32]{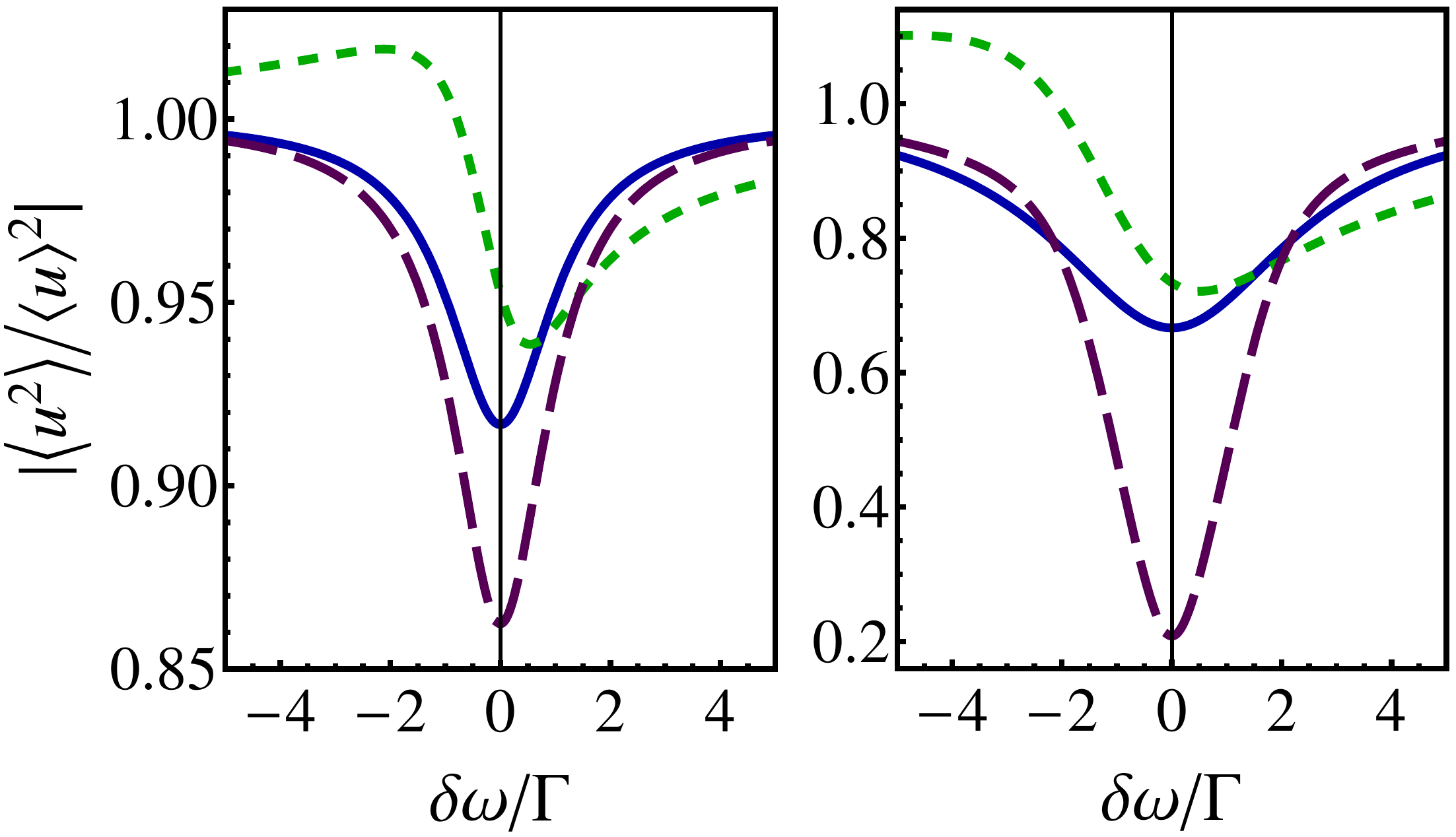}
\caption{(Color online) The scaled second moment of the complex amplitude of forced vibrations $u$ as a function of the frequency of the driving field, $\delta\omega=\omega_F-\omega_0$. The solid, short-dash and long-dash curves show the results for the Gaussian, Poisson, and telegraph noises, respectively. The left and right panels refer to the relative Gaussian noise intensities $D/\Gamma = 0.1$ and 1; note the different scales. For the Poisson noise on these panels, we took $g=1$ and the same intensities as the Gaussian noise, $\nu g^2/2=D$. For the telegraph noise, we chose $W_{12}=W_{21}=D$ and the variance $\xi_c^2=2D^2$.}
\label{fig:spectra}
\end{figure}

We now compare the effect of different types of frequency noise on the moments of the complex amplitude. We consider three common types of noise, broadband Gaussian and Poisson noises and telegraph noise; in the latter case, we choose  symmetric noise with $W_{12}=W_{21}$ and $\xi_2=-\xi_1$. In Fig.~\ref{fig:spectra}, we show the dependence of the second moment of the complex amplitude scaled by the squared mean amplitude on the frequency of the driving field. It is obtained from Eqs.~(\ref{eq:moments_delta_corr}) and (\ref{eq:telegraph_solution}). If there is no frequency noise, we have $\langle u^2\rangle/\langle u\rangle^2=1$. In the presence of noise the ratio $|\langle u^2\rangle/\langle u\rangle^2|$ can be smaller or larger than one. As seen from Fig.~\ref{fig:spectra}, this ratio displays resonant dependence on the field frequency. It most strongly differs from one near resonance, where $\omega_F=\omega_0$. As expected, the difference increases with the noise intensity.

\begin{figure}[h]
\centering
\includegraphics[scale=0.35]{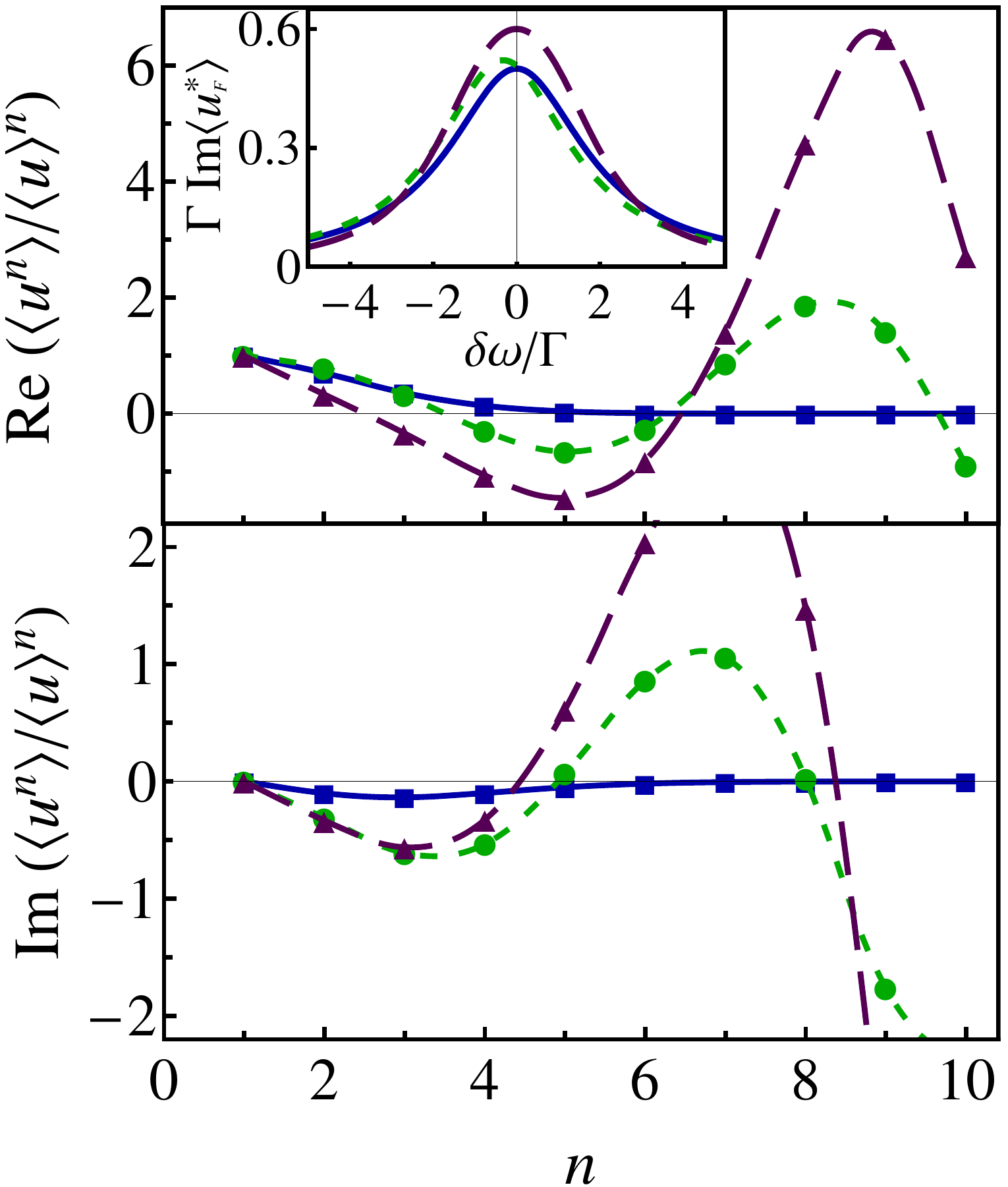}
\caption{(Color online) The real (upper panel) and imaginary (lower panel) parts of the normalized moments of the complex amplitude of  forced vibrations. The squares, circles, and triangles show the results for the Gaussian, Poisson, and telegraph noises, respectively; the lines are guides for the eye. The data refer to $\delta\omega/\Gamma=-1$ and the same noise parameters as in the right panel of Fig.~\ref{fig:spectra}. Inset: dependence of the scaled complex amplitude Im~$\langle u^*_F\rangle=(4\omega_F/F)~{\rm Im}\,\langle u^*\rangle$ on the driving field frequency for the different types of frequency noise; the plotted quantity gives the oscillator absorption spectrum. The coding of the curves and the noise parameters are the same as in the main figure.}
\label{fig:moments_off_resonance}
\end{figure}

For weak noise, $|\langle u^2\rangle/\langle u\rangle^2|$ is linear in the noise intensity, and  $|\langle u^2\rangle/\langle u\rangle^2|-1 \approx {\rm Re}\,\left[(\langle u^2\rangle -\langle u\rangle^2)/\langle u\rangle^2\right]$ is given by Eq.~(\ref{eq:2nd_moment_weak_noise}). Out of the three types of the noise discussed in this section, only for Poisson noise, we observe that $|\langle u^2\rangle/\langle u\rangle^2| >1$ in a certain frequency range. An interesting feature of this noise is that, even though $\langle\xi(t)\rangle = 0$, that is, $\omega_0$ is the average oscillator frequency, the frequency of the driving field where the field absorption is maximal , i.e., where Im~$u^*$ is maximal as a function of $\omega_F$, is located for $\omega_F-\omega_0 =-\nu(g - \sin g)$, see Eq.~(\ref{eq:Lorentzian_response}). This shift of the maximum of the absorption spectrum is seen in the inset in Fig.~\ref{fig:moments_off_resonance}.

\begin{figure}[h]
\centering
\includegraphics[scale=0.35]{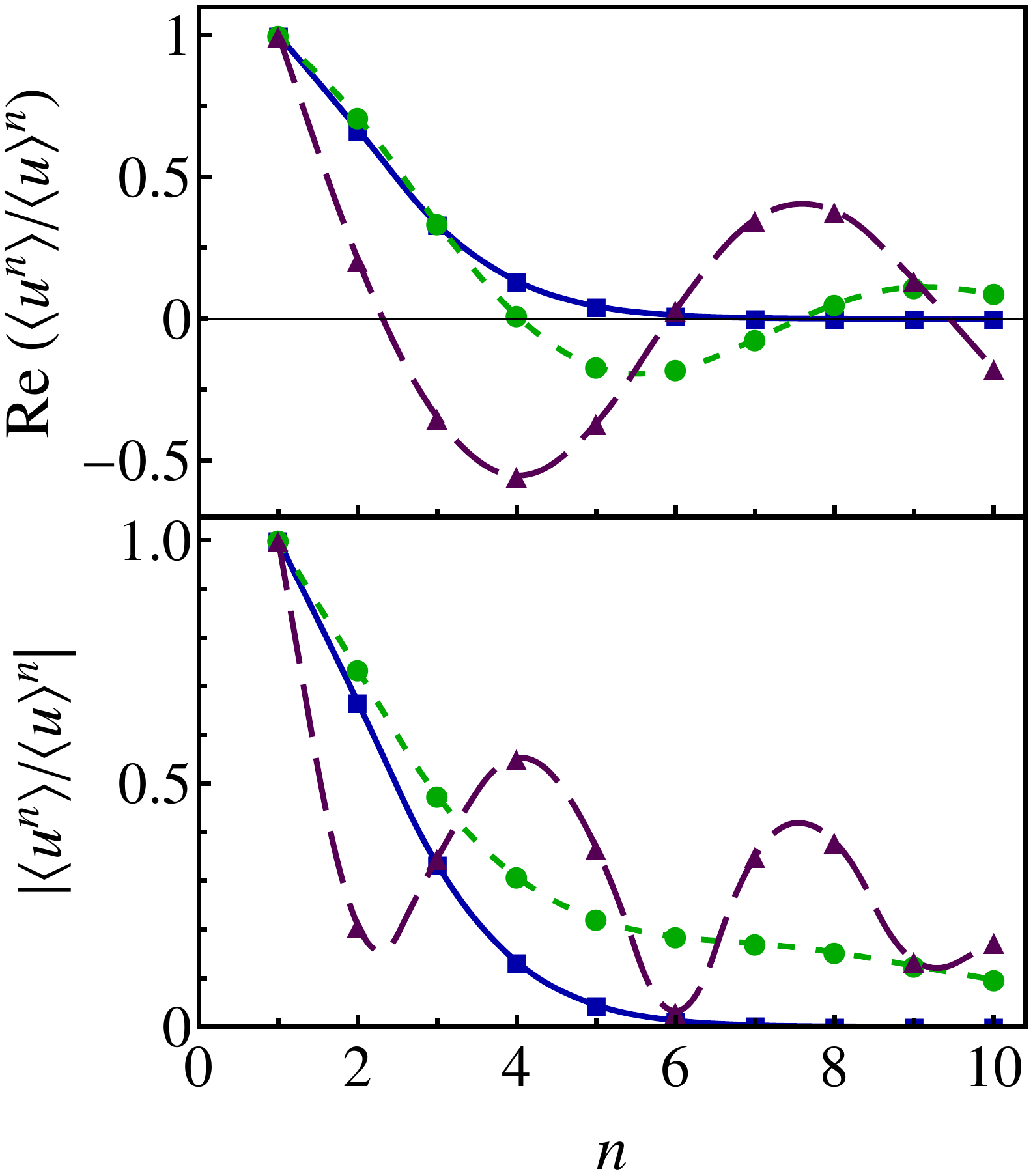}
\caption{(Color online) The real part and the absolute value of the normalized moments of the complex amplitude of forced vibrations at resonance, $\delta\omega=0$. The coding of the curves and the values of the noise parameters are the same as in Fig.~\ref{fig:moments_off_resonance}. }
\label{fig:resonant_moments}
\end{figure}

Figures \ref{fig:moments_off_resonance} and \ref{fig:resonant_moments} show higher-order moments of the complex amplitude for the two values of the frequency detuning, $\delta\omega/\Gamma = -1$ and  $\delta\omega=0$ (exact resonance).  The absorption spectra of the oscillator for the chosen noise parameters are shown in the inset of Fig.~\ref{fig:moments_off_resonance} . For Gaussian and Poisson frequency noises, these spectra are Lorentzian, and therefore the presence of the noise cannot be inferred from the spectrum. For a telegraph noise, the spectrum is non-Lorentzian, but clearly is close to a Lorentzian curve, even though the width of the spectral peak is increased by a factor $\sim 2$. At the same time, the moments of the complex amplitude unambiguously demonstrate the presence of frequency noise.

The dependence of the normalized moments on the order of the moment is nonmonotonic. It is very specific and markedly different for different types of noise. This is seen both in the real and imaginary parts of the moments and in their absolute values. We do not show the imaginary parts of the moments at exact resonance, since they are small there. It is seen also that, for Gaussian noise, the moments decrease more rapidly than for other noises we study, as it would be expected from Eq.~(\ref{eq:moments_delta_corr}). We note that the results for the moments of the complex amplitude of classical vibrations immediately apply to quantum vibrations, $\langle u^n\rangle^*=(\hbar/2\omega_0)^{n/2}\langle a^n\rangle$. 

Measurements of the moments and correlators of the complex amplitude can be done by standard homodyne detection in which the in-phase and quadrature components of the oscillator displacement are recorded as functions of time. This procedure is standard for classical oscillators.  For quantum oscillators,  measuring the moments $\langle a^n\rangle$ is simplified by the fact that operators $a^n(t)$ with different $n$ but the same time $t$ commute with each other. The moments can be immediately found, for example, from the Wigner distribution, which can be measured by means of Wigner tomography. \cite{Banaszek1996}  Experimental observation of the moments for microwave photons using a different procedure was reported recently. \cite{Eichler2011b}

\section{Conclusions}

 The results of this paper show that, for vibrational modes, the presence of frequency noise can be revealed and the statistics of the noise can be studied using the moments of the complex amplitude of forced vibrations $\langle u^n\rangle\propto \langle a^n\rangle^*$. The moments can be directly measured in the experiment. In the presence of frequency noise, they differ from the corresponding powers of the average complex amplitude $\langle u\rangle^n \propto (\langle a\rangle^n)^*$. The moments display a characteristic dependence on the frequency of the driving field and the moment number and are very sensitive to the noise statistics. This is illustrated using as examples Gaussian and Poisson noises with bandwidth that significantly exceeds the oscillator decay rate, as well as a telegraph noise.

Explicit expressions are obtained for the moments of the complex amplitude in the case of broadband noise with arbitrary statistics. A general formulation is developed for Markov noise, which reduces the problem of calculating the moments to a set of linear equations. Explicit results for the variance and the third cumulant of the complex amplitude are obtained also for an arbitrary noise provided the noise is weak; the third cumulant of the amplitude in this case is proportional to the third cumulant of the noise. Even for weak frequency noise, the proposed method allows revealing it irrespective of the intensity of additive noise in the oscillator. The results apply to both classical and quantum oscillators.

\begin{center}\bf{ACKNOWLEDGMENTS}\end{center}

This research was supported in part by DARPA through the DEFYS program. MID acknowledges also partial support from the NSF, Grant EMT/QIS 082985.

%\bibliographystyle{apsrev}
%\bibliographystyle{apsrev4-1}
%\bibliography{c:/Aaa/BibTex/md10}

%\end{document}
%

\end{document}